\documentclass[10pt]{amsart}

\usepackage{hyperref}
\usepackage{graphicx}
\usepackage{tikz}
\usepackage{amsthm}
\usepackage{multicol}

\usepackage{amssymb}
\usepackage{amsmath}

\let\newpf\proof \let\proof\relax

\def\DC{{\mathrm{DC}}}

\def\bm{\begin{matrix}}
\def\em{\end{matrix}}

\newcommand{\bt}{\begin{thm}}
\newcommand{\et}{\end{thm}}

\newcommand{\bl}{\begin{lemma}}
\newcommand{\el}{\end{lemma}}

\newcommand{\beq}{\begin{eqnarray}}
\newcommand{\eeq}{\end{eqnarray}}

\def\be{\begin{equation}}
\def\ee{\end{equation}}

\def\ba{{\begin{align}}}
\def\ea{{\end{align}}}

\def\0{{\mathbf 0}}

\newtheorem{thm}{Theorem}[section]

\newtheorem{lemma}[thm]{Lemma}

\newtheorem{obs}[thm]{Observation}
\theoremstyle{remark}
\newtheorem{rem}{Remark}[section]

\numberwithin{equation}{section}

\def \bn {\hfill \\ \smallskip\noindent}

\theoremstyle{definition}

\newtheorem{definition}{Definition}[section]
\def\proof{\bn {\bf Proof.} }

\def\note#1
{\marginpar

{\tiny $\leftarrow$
\par
\hfuzz=20pt \hbadness=9000 \hyphenpenalty=-100 \exhyphenpenalty=-100
\pretolerance=-1 \tolerance=9999 \doublehyphendemerits=-100000
\finalhyphendemerits=-100000 \baselineskip=6pt
#1}\hfuzz=1pt}

\renewcommand{\mod}{\operatorname{mod}}

\newcommand{\N}{{\mathbb N}}
\newcommand{\Q}{{\mathbb Q}}
\newcommand{\R}{{\mathbb R}}
\newcommand{\T}{{\mathbb T}}

\newcommand{\Z}{{\mathbb Z}}

\newcommand{\ap}{{\alpha^\prime}}
\newcommand{\sil}{\sigma(\lambda)}
\newcommand{\an}{\alpha^{(n)}}

\def\B0{{\bold{0}}}

\catcode`\@=12

\def\Empty{}
\newcommand\oplabel[1]{
  \def\OpArg{#1} \ifx \OpArg\Empty {} \else
  	\label{#1}
  \fi}

\newcommand{\comm}[1]{}
\newcommand{\comment}[1]{}

\begin{document}

\title[Spectral transition line for the extended Harper's model]{Spectral transition line for the extended Harper's model in the positive Lyapunov exponent regime}

\author{Fan Yang}

\address{Department of Mathematics, University of California, Irvine, California 92697-3875, USA}
\email{yangf13@uci.edu}

\thanks{}

\begin{abstract}
We study the spectral transition line of the extended Harper's model in the positive Lyapunov exponent regime. 
We show that both pure point spectrum and purely singular continuous spectrum occur for dense subsets of frequencies on the transition line.
\end{abstract}

\maketitle

\section{Introduction}  
Quasiperiodic Jacobi matrices arise naturally from the study of tight-binding electrons on a two-dimensional lattice exposed to a perpendicular magnetic field.
The most prominent example of such operators is the Harper's equation, mathematically known as the almost Mathieu operator (AMO), acting on $l^2(\Z)$, defined by (under a non-standard scaling),
\begin{align}\label{AMO}
(Hu)_n=\lambda(u_{n+1}+u_{n-1})+2\cos{2\pi (\theta+n\alpha)}u_n.
\end{align}

This paper considers a more general model that was introduced by D.J.Thouless in 1983 \cite{Thouless1983}, which includes the AMO as a special case. 
The Hamiltonian of the extended Harper's model (EHM), denoted by $H_{\lambda, \alpha, \theta}$, is defined as follow.
\begin{equation}\label{EHM}
(H_{\lambda,\alpha,\theta}u)_n=c_{\lambda}(\theta+n\alpha)u_{n+1}+\tilde{c}_{\lambda}(\theta+(n-1)\alpha)u_{n-1}+2\cos{2\pi (\theta+n\alpha)} u_n,
\end{equation}
acting on $l^2(\Z)$, in which
\begin{align*}
\left\lbrace
\begin{matrix}
&c_{\lambda}(\theta)=\lambda_1 e^{-2\pi i(\theta+\frac{\alpha}{2})}+\lambda_2 +\lambda_3 e^{2\pi i(\theta+\frac{\alpha}{2})}  \\
&\tilde{c}_{\lambda}(\theta)=\lambda_3 e^{-2\pi i(\theta+\frac{\alpha}{2})}+\lambda_2 +\lambda_1 e^{2\pi i(\theta+\frac{\alpha}{2})}
\end{matrix}
\right.
\end{align*}
We refer to $\alpha\in \T=[0,1]$ as the frequency and let $\beta(\alpha)$ (see (\ref{defbeta})) to be the upper exponent of exponential growth of continued fraction approximants. 
We also refer to $\theta\in \T$ as the phase and let $\lambda=(\lambda_1,\lambda_2,\lambda_3)$ be the coupling constant triple. 
The coupling constants are proportional to the probabilities an electron will hop to a corresponding neighbouring site. 
Without loss of generality, we assume $0\leq \lambda_{2} ~\mbox{,} ~0 \leq \lambda_{1} + \lambda_{3}$ and at least one of $\lambda_{1} \mbox{,} ~\lambda_{2} \mbox{,} ~\lambda_{3}$ to be positive. 
While the AMO ($\lambda_1=\lambda_3=0$) only takes nearest-neighbour hopping into account, the EHM includes next-nearest-neighbour hopping:

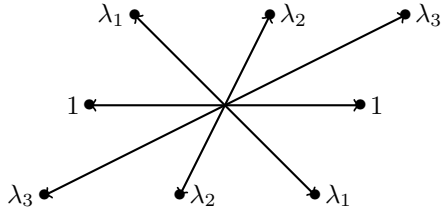
\begin{center}
\begin{tikzpicture}[thick, scale=0.6]
    \draw[->] (0, 0) -- (4, 2) node[right] {$\lambda_3$};
    \draw[->] (0, 0) -- (-4,-2) node[left] {$\lambda_3$};
    \draw[->] (0, 0) -- (3, 0) node[right] {$1$};
    \draw[->] (0, 0) -- (-3, 0) node[left] {$1$};
    \draw[->] (0, 0) -- (1, 2) node[right] {$\lambda_2$};
    \draw[->] (0, 0) -- (-1,-2) node[right] {$\lambda_2$};
    \draw[->] (0, 0) -- (-2, 2) node[left] {$\lambda_1$};
    \draw[->] (0, 0) -- (2, -2) node[right] {$\lambda_1$};
          \draw(4, 2) node [] {$\bullet$};
            \draw(-4, -2) node [] {$\bullet$};
              \draw(3, 0) node [] {$\bullet$};
                \draw(-3, 0) node [] {$\bullet$};
                  \draw(1, 2) node [] {$\bullet$};
                    \draw(-1, -2) node [] {$\bullet$};
                      \draw(-2, 2) node [] {$\bullet$};
                        \draw(2, -2) node [] {$\bullet$};
\end{tikzpicture}
\begin{description}
\item[Nearest-neighbour hopping] encoded in $\lambda_2$ and $1$,
\item[Next nearest-neighbour hopping] encoded in $\lambda_1$ and $\lambda_3$.
\end{description}
\end{center}

\

In the past few years, there have been several remarkable developments on obtaining arithmetic spectral transition for concrete quasiperiodic Schr\"odinger operators.
For the Maryland model, the spectral phase diagram was determined exactly for all $\alpha, \theta$ in \cite{JLmaryland}. 
For the AMO, the spectral transition conjecture in $\alpha$ \cite{amoconj94}: pure point spectrum (a.e. $\theta$) for $\beta(\alpha)<-\ln{\lambda}$ (under our non-standard scaling) and purely singular continuous spectrum for $\beta(\alpha)>-\ln{\lambda}$, was recently proved in \cite{AYZ}. 
Later, universal (reflective) hierarchical structure of eigenfunctions was established in the localization regime \cite{JL1,JL2} with also an arithmetic condition on $\theta$. 
Even more recently, the spectral transition line $\beta(\alpha)=-\ln{\lambda}$ was studied in \cite{AJZ}, where the authors showed that both pure point (for a.e. $\theta$) and purely singular continuous spectrum are dense phenomena.
 
The extended Harper's model, as a prime example of quasiperiodic (non-Schr\"odinger) Jacobi matrix, has also attracted great attention from both mathematics and physics (see e.g. \cite{DH, Hthou, Thouless1983}) literature. 
Under the classical duality map $\lambda=(\lambda_1, \lambda_2, \lambda_3) \rightarrow \sil=(\frac{\lambda_3}{\lambda_2}, \frac{1}{\lambda_2}, \frac{\lambda_1}{\lambda_2})$, the coupling constant space is naturally decomposed into the following three regions:
\begin{multicols}{2}
\begin{center}
\begin{tikzpicture}[thick, scale=0.6]
    \draw[->] (-10,-1) -- (-3,-1) node[below] {$\lambda_2$};
    \draw[->] (-10,-1) -- (-10,6) node[right] {$\lambda_1+\lambda_3$};
    \draw [ ] plot [smooth] coordinates { (- 7, 2) (-4, 5) };
    \draw [ ] plot [smooth] coordinates { (- 7, 2) (-7,-1) };
    \draw [ ] plot [smooth] coordinates { (-10, 2) (-7, 2) };

    \draw(-  4,   5) node [above] {$\lambda_1+\lambda_3=\lambda_2$};
    \draw(-  7,  -1) node [below] {$1$};
    \draw(- 10,   2) node [left]  {$1$};
    \draw(-9.7, 0.5) node [color=blue][right] {Region I};
    \draw(-  5.7,   1) node [color=blue][right] {Region II};
    \draw(-9.2, 3.6) node [color=blue][right] {Region III};

    \draw(-  7, 0.5) node [color=red][right] {$\mathrm{L}_{\mathrm{II}}$};
    \draw(-8.5,   2) node [color=red][above] {$\mathrm{L}_{\mathrm{I}}$};
    \draw(-3.5, 3.7) node [color=red][left] {$\mathrm{L}_{\mathrm{III}}$};

\end{tikzpicture}
\end{center}
\columnbreak
\

\begin{description}
\item[Region I] $0 \leq \lambda_{1}+\lambda_{3} \leq 1, 0 < \lambda_{2} \leq 1$,
\item[Region II] $0 \leq \lambda_{1}+\lambda_{3} \leq \lambda_{2}, 1 \leq \lambda_{2} $,
\item[Region III] $\max\{1,\lambda_{2}\} \leq \lambda_{1}+\lambda_{3}$, $\lambda_2>0$.
\end{description}
\

\ \ \ Let $\mathrm{I}^{\mathrm{o}}$, $\mathrm{II}^{\mathrm{o}}$, $\mathrm{III}^{\mathrm{o}}$ be the interiors, then,
\begin{itemize}
\item $\sigma(\mathrm{I}^\circ) = \mathrm{II}^\circ$, $\sigma(\mathrm{III}^\circ) = \sigma(\mathrm{III}^\circ)$
\item $\sigma(\mathrm{L}_\mathrm{I}) = \mathrm{L}_{\mathrm{III}}$ and $\sigma(\mathrm{L}_{\mathrm{II}})= \mathrm{L}_{\mathrm{II}}$,
\end{itemize}
\end{multicols}
As $\sigma$ bijectively maps $\mathrm{III} \cup \mathrm{L}_{\mathrm{II}}$ onto itself, the literature refers to $\mathrm{III} \cup \mathrm{L}_{\mathrm{II}}$ as the {\it self-dual regime}. We further divide $\mathrm{III}$ into $\mathrm{III}_{\lambda_1=\lambda_3}$ ({\it isotropic self-dual regime}) and $\mathrm{III}_{\lambda_1\neq \lambda_3}$ ({\it anisotropic self-dual regime}).

Recent developments on the spectral theory of the EHM include: 
pure point spectrum for Diophantine frequencies in the positive Lyapunov exponent regime $\mathrm{I}^{\mathrm{o}}$ \cite{JKS};
explicit formula for the Lyapunov exponent $L(\lambda)$ (see (\ref{ehmLE})) on the spectrum throughout all the three regions \cite{JMarx};
dry ten Martini problem for Diophantine frequencies in the self-dual regions \cite{dryten};
complete spectral decomposition for all $\alpha$ and a.e. $\theta$ in the zero Lyapunov exponent regiems \cite{AJM};
and arithmetic spectral transition in $\alpha$ in the positive Lyapunov exponent regime \cite{HJ}.
We could combine the results on spectral decomposition in the following theorem:
\begin{thm}
\
\begin{itemize}
\item \cite{AJM} For $\lambda \in \mathrm{II}^{\mathrm{o}} \cup \mathrm{III}^{\mathrm{o}}_{\lambda_1\neq \lambda_3}$, for all $\alpha$ and a.e. $\theta$, $H_{\lambda, \alpha, \theta}$ has purely absolutely continuous spectrum.

\item \cite{AJM, nopoint} For $\lambda \in \mathrm{III}^{\mathrm{o}}_{\lambda_1= \lambda_3} \cup \mathrm{L}_{\mathrm{I}} \cup \mathrm{L}_{\mathrm{II}} \cup \mathrm{L}_{\mathrm{III}}$, for all $\alpha$ and a.e. $\theta$, $H_{\lambda, \alpha, \theta}$ has purely singular continuous spectrum.

\item \cite{JKS,HJ} For $\lambda \in \mathrm{I}^{\mathrm{o}}$. 
If $\beta(\alpha)<L(\lambda)$, then for a.e. $\theta$, $H_{\lambda, \alpha, \theta}$ has pure point spectrum. 
If $\beta(\alpha)>L(\lambda)$, then for a.e. $\theta$, $H_{\lambda, \alpha, \theta}$ has purely singular continuous spectrum.
\end{itemize}
\end{thm} 

The missing link to a complete understanding of the whole picture for a.e. $\theta$ is the transition line: $\beta(\alpha)=L(\lambda)$ for $\lambda\in \mathrm{I}^{\mathrm{o}}$.
Filling this missing link, we show that both pure point and purely singular continuous spectra occur for dense subsets of frequencies on the transition line:
\begin{thm}\label{main}
For $\lambda\in \mathrm{I}^{\mathrm{o}}$, we have the following.
\begin{itemize}
\item for a dense subset of $\{\alpha:\ \beta(\alpha)=L(\lambda)\}$, $H_{\lambda, \alpha, \theta}$ has purely singular continuous spectrum for any $\theta$ in a full measure set $\Theta$.
\item for a dense subset of $\{\alpha:\ \beta(\alpha)=L(\lambda)\}$, $H_{\lambda, \alpha, \theta}$ has purely point spectrum for a.e. $\theta$.
\end{itemize}
\end{thm} 
\begin{rem}\label{Remark1.1}
The full measure set $\Theta$ appearing in the singular continuous part could be characterized as follows. We point out that $\Theta$ is independent of $\alpha$.
\begin{itemize}
\item if $\lambda_1\neq \lambda_3$ and $\lambda_1+\lambda_3\neq \lambda_2$, then $\Theta=\T$,
\item if $\lambda_1+\lambda_3=\lambda_2$, then $\Theta=\{\theta: \beta(2\theta)=0\}$,
\item if $\lambda_1=\lambda_3>\frac{\lambda_2}{2}$, then $\Theta=\{\theta: \beta(2\theta\pm \frac{1}{\pi}\arccos (-\frac{\lambda_2}{2\lambda_1}))=0\}$.
\end{itemize}
\end{rem}

Theorem \ref{main} is proved via perturbation arguments. 
The purely singular continuous part is proved by approaching a frequency $\alpha\in \{\alpha: \beta(\alpha)=L(\lambda)\}$ with $\alpha_n\in \{\alpha: \beta(\alpha)>L(\lambda)\}$, for which a Gordon-type argument could be applied. 
The pure point part follows from dual reducibility, that is obtained by approaching $\alpha$ with $\tilde{\alpha}_n\in \mathrm{DC}$, for which dual reducibility comes from localization (with eigenfunctions decaying at the rate of Lyapunov exponent) and a duality argument.

Thus the general strategy follows that of \cite{AJZ}, but we need to extend the aforementioned perturbation argument to our Jacobi setting. 
Such extensions, while often relatively straightforward in this non-singular case, are known to present (sometimes significant) technical difficulties for the singular one, which is the case here for a certain subset of $\lambda$.
In addition to those technical difficulties, an important difference lies in the fact that while for the AMO singular continuous spectrum for $\beta(\alpha)>-\ln{\lambda}$ holds for {\bf all} $\theta$, 
this is not even expected to be true for the EHM, as, for any $\alpha$, there are possibly some $\theta$ (depending on $\alpha$) with pure point spectrum like in the Maryland model. 
The possible dependence of the set of $\theta$ with singular continuous spectrum on $\alpha$ makes it hard to control the induction scheme.
This difficulty is overcome by our observation that the arithmetic exponent $\delta(\alpha, \theta)$ (see (\ref{defdelta})), coming from Gordon-type argument, coincides with $\beta(\alpha)$ for an {\bf $\alpha$-independent full measure set of $\theta$}, see Remark \ref{Remark1.1}.
This purely number theoretical fact is proved in Lemma \ref{delta=beta}, which we believe to be of independent interest. 
This, in particular, enables us to conduct our perturbation and induction scheme in the singular case (when $c_{\lambda}(\cdot)$ has zeros).
Another fact of independent interest is Theorem \ref{asymthm}. It gives the asymptotic behavior of eigenfunctions in the positive Lyapunov exponent regime for Diophantine frequencies. 
On one hand, this result adds to the growing collection of exact characterizations of quasiperiodic eigenfunctions.
On the other hand, the exact decay rate is crucial to our proof of the pure point spectrum part.
We point out that although localization has been established in the same regime \cite{JKS}, the asymptotic behavior of eigenfunctions was not proved there. 

We organize the paper as follows: 
in section 2 we present some preliminaries, in section 3 we show two key lemmas (Lemma \ref{delta=beta} and Theorem \ref{asymthm}), 
and we complete the proofs of purely singular continuous and pure point part in sections 4 and 5 respectively.

\section{Preliminaries}
For $x\in \R$, let $\|x\|=\mathrm{dist}(x, \Z)$.
\subsection{Cocycles and Lyapunov exponent}

The eigenvalue equation $H_{\lambda,\alpha,\theta}u=Eu$ can be rewritten via the following equation
 \begin{align*}
\left (\begin{matrix}  u_{n+1} \\ u_{n} \end{matrix} \right )= A^{E}_{\lambda, \alpha}(\theta+n\alpha)\left (\begin{matrix}  u_{n} \\  u_{n-1}  \end{matrix} \right )
 \end{align*} 
where
\begin{align*}
A^{E}_{\lambda, \alpha}(\theta)=\frac{1}{c_{\lambda}(\theta)}\left (\begin{matrix}  E-2\cos{2\pi \theta}   &  -\tilde{c}_{\lambda}(\theta-\alpha) \\ c_{\lambda}(\theta)  &   0 \end{matrix} \right )
\end{align*}
is called the {transfer matrix}.
We call $(\alpha, A^{E}_{\lambda, \alpha})$ the associated {cocycle} of $H_{\lambda, \alpha, \theta}$ at energy $E$.
 
Let 
\begin{align*}
A^E_{\lambda, \alpha, n}(\theta)=A^E_{\lambda, \alpha}(\theta+(n-1)\alpha)A^E_{\lambda, \alpha}(\theta+(n-2)\alpha)\cdots A^E_{\lambda, \alpha}(\theta)
\end{align*}
be the {n-step transfer matrix}

We can then define the {Lyapunov exponent} of $H_{\lambda, \alpha, \cdot}$ at energy $E$ to be:
\begin{align}\label{defLE}
L(\lambda, \alpha, E)=\lim_{n\rightarrow\infty}\frac{1}{n}\int_{\T}\ln{\|A^E_{\lambda, \alpha, n}(\theta)\|}\ \mathrm{d} \theta.
\end{align}

One remarkable feature of the extended Harper's model is that its Lyapunov exponent on the spectrum could be computed explicitly.
\begin{thm}\cite{JMarx}\label{ehmLE}
For $E\in \Sigma_{\lambda, \alpha}$, we have,
\begin{align}
L(\lambda, \alpha, E)
\left\lbrace 
\begin{matrix}
\equiv L(\lambda)=\ln{ \frac{1+\sqrt{1-4\lambda_1 \lambda_3}}{\max{(\lambda_1+\lambda_3,\lambda_2)} +\sqrt{\max{(\lambda_1+\lambda_3,\lambda_2)}^2-4\lambda_1\lambda_3}}}>0,\ \ &\lambda\in \mathrm{I}^{\mathrm{o}},\\
0,\ \ &\mathrm{otherwise}.
\end{matrix}
\right.
\end{align}  
\end{thm}

In order to employ reducibility methods, we introduce {normalized transfer matrix}:
\begin{align*}
\tilde{A}^E_{\lambda, \alpha}(\theta)
=\frac{1}{\sqrt{|c|_{\lambda}(\theta)|c|_{\lambda}(\theta-\alpha)}}
\left (\begin{matrix}  
E-2\cos{2\pi \theta}   &  -|c|_{\lambda}(\theta-\alpha) 
\\ |c|_{\lambda}(\theta)  &   0 
\end{matrix} \right ),
\end{align*} 
where $|c|_{\lambda}(\theta)=\sqrt{c_{\lambda}(\theta)\tilde{c}_{\lambda}(\theta)}$. 
$(\alpha, \tilde{A}^E_{\lambda, \alpha})$ will be called the normalized cocycle.

As was introduced in \cite{dryten}, the advantage of $(\alpha, \tilde{A}^E_{\lambda, \alpha})$ over $(\alpha, A^E_{\lambda, \alpha})$ is the that the normalized matrix $\tilde{A}^E_{\lambda, \alpha}(\theta)$ is homotopic to identity in $C^{0}(\T, \mathrm{SL}(2,\R))$.
Thus making it a nicer object to apply techniques from reducibility.

\subsection{Aubry duality of extended Harper's model}
The spectrum $\Sigma_{\lambda, \alpha}$ of $H_{\lambda, \alpha, \cdot}$ is related to the spectrum $\Sigma_{\sil, \alpha}$ of $H_{\sil, \alpha, \cdot}$ in the following way
\begin{align*}
\Sigma_{\lambda, \alpha}=\lambda_2 \Sigma_{\sil, \alpha}
\end{align*}
by Aubry duality.

\subsection{Reducibility, rotation number}
Given two cocycles $(\alpha, A^{(1)})$ and $(\alpha, A^{(2)})$, we say they are $C^k$ ($k=\omega$ or $\infty$) to each other if there exists $B\in C^k(\T, \mathrm{PSL}(2,\R))$ such that $B(x+\alpha)A^{(1)}(x)B^{-1}(x)=A^{(2)}(x)$.
If $(\alpha, A)$ is $C^k$ conjugate to a constant cocycle, then we say $(\alpha, A)$ is $C^k$ reducible.

Let 
\begin{align*}
R_{x}=
\left(
\begin{matrix}
\cos{2\pi x} \ \ &-\sin{2\pi x}\\
\sin{2\pi x} \ \ &\cos{2\pi x}
\end{matrix}
\right).
\end{align*}
Any $A\in C^0(\T, \mathrm{PSL}(2,\R))$ is homotopic to $x\rightarrow R_{\frac{k}{2}x}$ for some $k\in \Z$ called the degree of $A$, denoted by $\deg{A}=k$.

Assume now that $A\in C^0(\T, \mathrm{SL}(2,\R))$ is homotopic to identity.
Then there exists $\phi:\R/\Z \times \R/\Z \to \R$ and $v:\R/\Z \times \R/\Z \rightarrow \R^+$ such that
\begin{align*}
A(x) 
\left(
\begin{matrix}
\cos 2 \pi y \\ 
\sin 2 \pi y 
\end{matrix} 
\right)
=v(x,y)
\left(
\begin{matrix} 
\cos 2 \pi (y+\phi(x,y)) \\
\sin 2 \pi (y+\phi(x,y)) 
\end{matrix} 
\right).
\end{align*}
The function $\phi$ is called a lift of $A$.  
Let $\mu$ be any probability on $\R/\Z \times \R/\Z$ which is invariant under the continuous
map $T:(x,y) \mapsto (x+\alpha,y+\phi(x,y))$, projecting over Lebesgue
measure on the first coordinate.  
Then the number
\begin{align*}
\rho(\alpha,A)=\int \phi\ d\mu \mod \Z
\end{align*}
is independent of the choices of $\phi$ and $\mu$, and is called the
{fibered rotation number} of
$(\alpha,A)$ \cite{Herman, JohnsonMoser}.

Rotation number plays a fundamental role in the reducibility theory. 
Readers can consult Theorem 1.5 of \cite{YouZhou} and the discussions therein. Particularly, we have the following: 

\begin{thm}\label{KAM}\cite{AFK, HouYou, YouZhou}
Let $(\alpha,A)\in \R \setminus \Q \times C_h^{\omega}(\T, \mathrm{SL}(2,\R))$ with $h>h^{\prime}>0$, $R\in \mathrm{SL}(2,\R)$. 
Then for every $\tau >1$, $\gamma>0$, if $\rho(\alpha,A)\in \mathrm{DC}_{\alpha}(\tau, \gamma)$, then there exists $\varepsilon=\varepsilon(\tau,\gamma,h-h^{\prime})$, such that if $\|A(\theta)-R\|_h<\varepsilon(\tau,\gamma,h-h^{\prime})$, then there exists $B\in C_{h^{\prime}}^{\omega}(\T, \mathrm{SL}(2,\R))$, $\phi\in C_{h^{\prime}}^{\omega}(\T,\R)$, such that
\begin{equation}
B(\theta+\alpha)A(\theta)B(\theta)^{-1}=R_{\phi(\theta)}
\end{equation}
Moreover, we have the following estimates
\begin{enumerate}
\item $\|B-\mathrm{Id}\|_{h^{\prime}}\leq \|A(\theta)-R\|_h^{\frac{1}{2}},$
\item $\|\phi(\theta)-\hat{\phi}(0)\|_{h^{\prime}}\leq 2\|A(\theta)-R\|_h.$
\end{enumerate}
\end{thm}

\subsection{Continued fraction}\label{seccontinued}
Another important tool is the continued fraction expansion of irrational numbers.

Let $\alpha\in \T\setminus \Q$, $\alpha$ has the following unique expression with $a_n\in \N$: 
\begin{align}\label{defcontinuefractionan}
\alpha=\frac{1}{a_1+\frac{1}{a_2+\frac{1}{a_3+\cdots}}}.
\end{align}
Let 
\begin{align}\label{defpnqn}
\frac{p_n}{q_n}=\frac{1}{a_1+\frac{1}{a_2+\frac{1}{\cdots+\frac{1}{a_n}}}}
\end{align}
be the continued fraction approximants of $\alpha$. 
Let
\begin{align}\label{defbeta}
\beta(\alpha)=\limsup_{n\rightarrow\infty}\frac{\ln{q_{n+1}}}{q_n}.
\end{align}
$\beta(\alpha)$ being large means $\alpha$ can be approximated very well by a sequence of rational numbers.
Let us mention that $\{\alpha: \beta(\alpha)=0\}$ is a full measure set.
It is clear from (\ref{defpnqn}) and (\ref{defbeta}) that
\begin{align}\label{beta=limsupa}
\beta(\alpha)=\limsup_{n\rightarrow\infty}\frac{\ln{a_{n+1}}}{q_n}.
\end{align}

The following properties about continued fraction expansion are well-known:
\begin{align}\label{qnqn+1}
\frac{1}{2q_{n+1}}\leq \|q_n\alpha\|_{\T}\leq \frac{1}{q_{n+1}}.
\end{align}
\begin{align}\label{alpha-pnqn}
|\alpha-\frac{p_n}{q_n}|\leq \frac{1}{q_{n}q_{n+1}}.
\end{align}
For any $q_n\leq |k|<q_{n+1}$,
\begin{align}\label{qnkqn+1}
\|q_n\alpha\|_{\T}\leq \|k\alpha\|_{\T}.
\end{align}
Combining definition of $\beta(\alpha)$ (\ref{defbeta}) with (\ref{qnkqn+1}), we have:
If $\beta(\alpha)=0$, then for any $\epsilon>0$, for $|k|$ large, the following inequality holds:
\begin{align}\label{beta=0epsilon}
\|k\alpha\|_{\T}>e^{-\epsilon |k|}.
\end{align}

Now we introduce the Diophantine condition,
for any $\tau>1$, $\gamma>0$, let 
\begin{align}\label{defDCtaugamma}
\mathrm{DC}(\tau, \gamma)=\{\alpha: \|k\alpha\|_{\T}\geq \frac{\gamma}{(|k|+1)^{\tau}}\}.
\end{align}
It is clear that for any $\tau>1$, $\mathrm{DC}(\tau):=\cup_{\gamma>0}\mathrm{DC}(\tau, \gamma)$ is a full measure set. 
Let us denote $\mathrm{DC}=\cup_{\tau>1}\cup_{\gamma>0}\mathrm{DC}(\tau, \gamma)$.
It can be easily seen that if $a_n(\alpha)\equiv 1$ for $n$ large enough, then $\alpha\in \mathrm{DC}$. This fact will be used several times in section 5. 

Now for a fixed $\alpha$, we introduce Diophantine condition with respect to $\alpha$. For any $\tau>1$, $\gamma>0$, let
\begin{align}\label{defDioalpha}
\mathrm{DC}_{\alpha}(\tau, \gamma)=\{\theta: \|2\theta-k\alpha\|_{\T}\geq \frac{\gamma}{(|k|+1)^{\tau}}\}.
\end{align}
It is also clear that for any $\tau>1$, $\mathrm{DC}_{\alpha}(\tau):=\cup_{\gamma>0}\mathrm{DC}_{\alpha}(\tau, \gamma)$ is a full measure set.

\section{Key lemmas}
\subsection{Zeros of $c_{\lambda}(\theta)$}\label{Zeroc(theta)}
\

One significant difference between the extended Harper's model and Schr\"odinger operators is the presence of zeros of the off-diagonal sampling function.
\begin{obs}\label{zeros}(see e.g. \cite{AJM})
\begin{itemize}
\item when $\lambda_1\neq \lambda_3$ and $\lambda_1+\lambda_3\neq \lambda_2$, $c_{\lambda}(\theta)$ has no zeros on $\T$,
\item when $\lambda_1+\lambda_3=\lambda_2$, $c_{\lambda}(\theta)$ has a single zero point $\theta_1=\frac{1}{2}-\frac{\alpha}{2}$,
\item when $\lambda_1= \lambda_3>\frac{\lambda_2}{2}$, $c_{\lambda}(\theta)$ has two different zeros $\theta_1=\frac{1}{2\pi}\arccos(-\frac{\lambda_2}{2\lambda_1})-\frac{\alpha}{2}$ and $\theta_2=-\frac{1}{2\pi}\arccos(-\frac{\lambda_2}{2\lambda_1})-\frac{\alpha}{2}$.
\end{itemize}
\end{obs}
Now let $\theta_1,...,\theta_m$ be the zeros of $c_{\lambda}(\theta)$, define the following exponent $\delta$:
\begin{align}\label{defdelta}
\delta(\alpha, \theta)=\limsup_{n\rightarrow \infty}\frac{\sum_{k=1}^m \ln{\|q_n(\theta-\theta_k)\|_{\T}+\ln{q_{n+1}}}}{q_n}\leq \beta(\alpha).
\end{align}
Obviously, when $\lambda_1\neq \lambda_3$ and $\lambda_1+\lambda_3\neq \lambda_2$, $\delta(\alpha, \theta)\equiv \beta(\alpha)$. 
While when $c_{\lambda}(\theta)$ has zeros, we have the following relation between $\delta(\alpha, \theta)$ and $\beta(\alpha)$.

\begin{lemma}\label{delta=beta}
Let $\beta(\alpha)>0$.
For $\theta\in \Theta$, we have $\delta(\alpha, \theta)=\beta(\alpha)$. More precisely, we have,
\begin{itemize}
\item when $\lambda_1+\lambda_3=\lambda_2$, if $\theta\in \{\theta: \beta(2\theta)=0\}$, then $\delta(\alpha, \theta)=\beta(\alpha)$.
\item when $\lambda_1=\lambda_3>\frac{\lambda_2}{2}$, if $\theta\in \{\theta: \beta(2\theta\pm \frac{1}{\pi}\arccos (-\frac{\lambda_2}{2\lambda_1}))=0\}$, then $\delta(\alpha, \theta)=\beta(\alpha)$.
\end{itemize}
\end{lemma}
\subsection*{Proof of Lemma \ref{delta=beta}}
We will prove the case when $\lambda_1+\lambda_3=\lambda_2$ and $\beta(2\theta)=0$. The other one can be proved similarly.
For any small $\epsilon>0$, take a subsequence $\{\frac{p_{n_k}}{q_{n_k}}\}$ such that $\frac{\ln{q_{n_k+1}}}{q_{n_k}}>\beta(\alpha)-\epsilon$.

Then if $\|q_{n_k}(\theta-\frac{1}{2}+\frac{\alpha}{2})\|_{\T}\in [\frac{1}{4}, \frac{1}{2}]$, we have
\begin{align}\label{case1case1}
\limsup_{k\rightarrow\infty}\frac{\ln \|q_{n_k}(\theta-\frac{1}{2}+\frac{\alpha}{2})\|_{\T}+\ln{q_{n_k+1}}}{q_{n_k}}
\geq \limsup_{k\rightarrow\infty}\frac{\ln (1/4)+\ln{q_{n_k+1}}}{q_{n_k}}\geq \beta(\alpha)-\epsilon.
\end{align}

On the other hand, if $\|q_{n_k}(\theta-\frac{1}{2}+\frac{\alpha}{2})\|_{\T}\in [0, \frac{1}{4})$, we have 
\begin{align}
\|q_{n_k}(\theta-\frac{1}{2}+\frac{\alpha}{2})\|_{\T}
=&\frac{1}{2}\|2q_{n_k}(\theta-\frac{1}{2}+\frac{\alpha}{2})\|_{\T}\notag\\
=&\frac{1}{2}\|q_{n_k}(2\theta)+q_{n_k}\alpha\|_{\T}\notag\\
\geq &\frac{1}{2}(\|q_{n_k}(2\theta)\|_{\T}-\|q_{n_k}\alpha\|_{\T})\notag\\
\geq &\frac{1}{2}(\|q_{n_k}(2\theta)\|_{\T}-\frac{1}{q_{n_k+1}})\label{case1useqnqn+1}\\
\geq &\frac{1}{2}(e^{-\epsilon q_{n_k}}-e^{-(\beta-\epsilon)q_{n_k}})\label{case1usebeta=0}\\
> &e^{-2\epsilon q_{n_k}},\notag
\end{align}
where we applied (\ref{qnqn+1}) in (\ref{case1useqnqn+1}), and (\ref{beta=0epsilon}) in (\ref{case1usebeta=0}).
Hence 
\begin{align}\label{case1case2}
\limsup_{k\rightarrow\infty}\frac{\ln \|q_{n_k}(\theta-\frac{1}{2}+\frac{\alpha}{2})\|_{\T}+\ln{q_{n_k+1}}}{q_{n_k}}
\geq \limsup_{k\rightarrow\infty} \frac{-2\epsilon q_{n_k}+(\beta(\alpha)-\epsilon)q_{n_k}}{q_{n_k}}\geq \beta(\alpha)-3\epsilon.
\end{align}
Therefore, combining (\ref{case1case1}) with (\ref{case1case2}), we have $\delta(\alpha, \theta)\geq \beta(\alpha)-3\epsilon$, for any $\epsilon>0$. This implies $\delta(\alpha,\theta)=\beta(\alpha)$ for any $\theta\in \{\theta: \beta(2\theta)=0\}$.
$\hfill{} \Box$

\subsection{Asymptotic of eigenfunctions in region $\mathrm{I}^{\mathrm{o}}$}\label{asym}
\

\begin{thm}\label{asymthm}
Let $\lambda\in \mathrm{I}^{\mathrm{o}}$, $\alpha\in \mathrm{DC}$ and $\theta\in \mathrm{DC}_{\alpha}(\tau)$ for some $\tau>1$. Then for any eigenvalue $E$ of $H_{\lambda, \alpha, \theta}$, the corresponding eigenfunction $\phi_E$ satisfies:
\begin{align*}
\lim_{|n|\rightarrow\infty}\frac{\ln{(\phi_E^2(n)+\phi_E^2(n+1))}}{2|n|}=-L(\lambda).
\end{align*}
\end{thm}
\subsection*{Proof of Theorem \ref{asymthm}}
\

Note that since $
\lim_{|n|\rightarrow\infty}\frac{\ln{(\phi_E^2(n)+\phi_E^2(n+1))}}{2|n|}\geq -L(\lambda)
$
is obvious, it suffices to prove the other direction.

The proof will be based on \cite{JKS}. 
Let $H_{\lambda, \alpha, \theta}[x_1, x_2]$ be the restriction of $H_{\lambda, \alpha, \theta}$ to the interval $[x_1, x_2]$ with zero boundary condition. For $x, y\in [x_1, x_2]$, let $G_{[x_1, x_2]}(x, y)=(H_{\lambda, \alpha, \theta}[x_1, x_2]-E)^{-1}(x, y)$. Let us introduce the following definition:
\begin{definition}\label{regularpoint}
Let $m>0$ and $k\in \N$. A point $y\in \Z$ will be called $(m, k)$-regular if there exists an interval $[x_1, x_2]$ with $x_2=x_1+k-1$ containing $y$ such that
\begin{align*}
|G_{[x_1, x_2]}(y, x_i)|<e^{-m|y-x_i|}
\end{align*}
and $\mathrm{dist}(y, x_i)\geq \frac{1}{9}k$. Otherwise $y$ will be called $(m, k)$-singular.
\end{definition}
\begin{rem}
We point out that this definition is a little different from that in \cite{JKS}, 
where a regular point is to satisfy $|G_{[x_1, x_2]}(y, x_i)|<e^{-\frac{mk}{9}}$. 
However, following exactly the same proof, the following lemma about the repulsion between two singular points is still true.
\end{rem}
\begin{lemma}\label{repulsion}\cite{JKS}
Let $\lambda, \alpha, \theta$ be as in Theorem \ref{asymthm}. 
Then if $\epsilon\in (0, \frac{L(\lambda)}{3})$ and $1<a<2$, there exists $K(\alpha, \theta, \epsilon, a, E)\in \N$: such that for any $k\geq K(\alpha, \theta, \epsilon, a, E)$, if $y$ is $(L(\lambda)-\epsilon, k)$-singular and $|y|>\frac{3}{4}k$, then $|y|>(k-2)^a$.
\end{lemma}
Now in order to prove Theorem \ref{asymthm}, we first fix $1<a_1<a<2$. 
Let $E$ be an eigenvalue of $H_{\lambda, \alpha, \theta}$. Since $\phi_E\in l^2$, we can normalize it to have an a priori bound
\begin{align*}
|\phi_E(n)|\leq 1.
\end{align*}
Now take $|x|$ large enough, and take integer $k$ such that $(k-1)^{a_1}<|x|\leq k^{a_1}$. 

By Lemma \ref{repulsion}, for any $|y|\in [\frac{3}{4}k, (k-2)^a]$, we know $y$ is $(L(\lambda)-\epsilon, k)$-regular. 
This means for any such $y$, there exists an interval $I(y):=[x_1, x_2]\ni y$ with length $k$ such that $|G_{I(y)}(y, x_i)|<e^{-m|y-x_i|}$ and $|y-x_i|\geq \frac{k}{9}$. 
We denote the boundary of $I(y)$ by $\partial I(y):=\{x_1, x_2\}$, and for each $z\in \partial I(y)$, let $z^\prime$ be the neighbor of $z$ (i.e. $|z-z^\prime|=1$) not belong to $I(y)$.

Now starting with $x$, we could expand $\phi_E(x)$ using the following Green's function expansion:
\begin{align}\label{Greenexpansion}
\phi_E(x)=G_{I(x)}(x_1, x)\phi_E(x_1-1)+G_{I(y)}(x_2, y)\phi_E(x_2+1).
\end{align}
Since clearly $|x_1-1|, |x_2+1|$ are still inside $[\frac{3}{4}k, (k-2)^a]$, we could expand the term $\phi_E(x_1-1)$  using (\ref{Greenexpansion}) with $I(x_1-1)$, and $\phi_E(x_2+1)$ with $I(x_2+1)$. 
We continue to expand the terms of the form $\phi_E(z)$ in the same fashion until we arrive at such a $z$ 
that either $|z|<\frac{3}{4}k$ or the number of $G_I$ terms in the product becomes $9k^{a_1-1}$, whichever comes first. 
We then obtain an expression of the form
\begin{align}\label{Greenexpansionmany}
\phi_E(x)=\sum_{s; z_{i+1}\in \partial I(z_i^\prime)}G_{I(x)}(x, z_1) G_{I(z_1^\prime)}(z_1^\prime, z_2)\cdots G_{I(z_s^\prime)}(z_s^\prime, z_{s+1})\phi_E(z_{s+1}^\prime),
\end{align}
where in each term of the summation we have $|z_i|>\frac{3}{4}k$, $i=1,...,s$, and either $|z_{s+1}^\prime|<\frac{3}{4}k$, $s\leq 9k^{a_1-1}$, or $s+1=9k^{a_1-1}$.

If $|z_{s+1}^\prime|<\frac{3}{4}k$, then by the definition of $(L(\lambda)-\epsilon, k)$-regularity, we have,
\begin{align}\label{expand1}
\notag &|G_{I(x)}(x, z_1) G_{I(z_1^\prime)}(z_1^\prime, z_2)\cdots G_{I(z_s^\prime)}(z_s^\prime, z_{s+1})\phi_E(z_{s+1}^\prime)|\\
\notag <&e^{-(L(\lambda)-\epsilon)(|z_1-x|+\sum_{i=1}^s |z_{i+1}-z_i^\prime|)}\\
\notag \leq &e^{-(L(\lambda)-\epsilon)(|z_1-x|+\sum_{i=1}^s (|z_{i+1}-z_i|-1))}\\
\notag \leq &e^{-(L(\lambda)-\epsilon)(|x|-\frac{3}{4}k-9k^{a_1-1})}\\
\leq &e^{-(L(\lambda)-2\epsilon)|x|}.
\end{align}

If $s+1=9k^{a_1-1}$, then by the definition of $(L(\lambda)-\epsilon, k)$-regularity and use the fact that $|x-z_1|\geq \frac{k}{9}$ and $|z_{i+1}-z_i^\prime|\geq \frac{k}{9}$, we have,
\begin{align}\label{expand2}
\notag &|G_{I(x)}(x, z_1) G_{I(z_1^\prime)}(z_1^\prime, z_2)\cdots G_{I(z_s^\prime)}(z_s^\prime, z_{s+1})\phi_E(z_{s+1}^\prime)|\\
\notag <&e^{-(L(\lambda)-\epsilon)(|z_1-x|+\sum_{i=1}^s |z_{i+1}-z_i^\prime|)}\\
\notag <&e^{-(L(\lambda)-\epsilon)\frac{k}{9} 9k^{a_1-1}}\\
\leq &e^{-(L(\lambda)-\epsilon)|x|}.
\end{align}
Finally, we observe that the total number of terms in (\ref{Greenexpansionmany}) could be bounded above by $2^{9k^{a_1-1}}$, thus combining (\ref{Greenexpansionmany}), (\ref{expand1}) with (\ref{expand2}) we have,
\begin{align*}
|\phi_E(x)|\leq 2^{9k^{a_1-1}} e^{-(L(\lambda)-2\epsilon)|x|}<e^{-(L(\lambda)-3\epsilon)|x|}.
\end{align*}
$\hfill{} \Box$

\section{Singular continuous spectrum}
Throughout this section, we will assume $\theta\in \Theta$.

The following definition of $(C, N)$-badness was first introduced in \cite{AYZ}.

\begin{definition}\label{CNbad}
Given any $C>0$, $N\in \N$. If for any $E\in \Sigma_{\lambda, \alpha}$, and for any normalized solution $u$, in the sense that $|u(0)|^2+|u(-1)|^2=1$, of $H_{\lambda, \alpha, \theta}u=Eu$, we have
$\sum_{|k|\leq N} |u(k)|^2 \geq C^2$.
Then we say $(\lambda,\alpha, \theta)$ is $(C,N)$-bad.
\end{definition} 
 
The singular continuous part of our main theorem follows naturally from the next lemma:                                                           
\begin{lemma}\label{CNpsc}
Let $\lambda\in \mathrm{I}^{\mathrm{o}}$, if for any $C>0$, there exists $N\in \N$ such that $(\lambda, \alpha, \theta)$ is $(C,N)$-bad. Then $H_{\lambda, \alpha, \theta}$ has pure singular continuous spectrum.
\end{lemma}

The following two lemmas will be crucial to our proof.

\begin{lemma}\label{existCNbad}
Let $\lambda\in \mathrm{I}^{\mathrm{o}}$ be such that $L(\lambda)<\beta(\alpha)$, then 
for any $C>0$, there exists $N\in \N$ such that $(\lambda, \alpha, \theta)$ is $(C, N)$-bad for any $\theta\in \Theta$.
\end{lemma}

\subsection*{Proof of Lemma \ref{existCNbad}}
Following a Gordon-type argument (see Theorem 1.3 of \cite{HJ}), we can show that for any $C>0$, there exists $N\in \N$ such that $(\lambda, \alpha, \theta)$ is $(C, N)$-bad if $L(\lambda)<\delta(\alpha, \theta)$.
Lemma \ref{existCNbad} then follows immediately from Lemma \ref{delta=beta}.
$\hfill{} \Box$

\begin{lemma}\label{CNCprimeN}
If $(\lambda,\alpha, \theta)$ is $(C,N)$-bad, then for any $0<C^{\prime} < C$, there exists $\epsilon(\lambda,\alpha,\theta, C-C^{\prime},N)>0$, such that $(\lambda,\alpha^{\prime},\theta)$ is $(C^{\prime},N)$-bad as long as $|\alpha-\alpha^{\prime}|<\epsilon$. 
\end{lemma}

\subsection*{Proof of Lemma \ref{CNCprimeN}} 
By H\"older continuity of the spectrum in Hausdorff topology,
for any $E^{\prime} \in \Sigma_{\lambda, \alpha^\prime}$, there exists $E \in \Sigma_{\lambda,\alpha}$ 
with $|E-E^{\prime}|<C(\lambda)|\alpha^{\prime}-\alpha|^{\frac{1}{2}}<C(\lambda)\epsilon^{\frac{1}{2}}$.
Thus for any $|m|\leq N$, we have
\begin{align*}
      & |(E-2\cos 2\pi (\theta+m\alpha))-(E^{\prime}-2\cos 2\pi (\theta+m\alpha^{\prime}))| \\
\leq& C(\lambda)\epsilon^{\frac{1}{2}}+C(N)|\alpha-\alpha^{\prime}| \\
\leq &C(\lambda, N)\epsilon^{\frac{1}{2}}.
\end{align*}
Clearly, we also have
\begin{align*}
      &|c_{\lambda}(\theta+m \alpha)-c_{\lambda}(\theta+m\alpha^{\prime})| \\
\leq &|\lambda_1e^{-2\pi i(\theta+m\alpha+\frac{\alpha}{2})}-\lambda_1e^{-2\pi i(\theta+m\alpha^\prime+\frac{\alpha^{\prime}}{2})}| +|\lambda_3e^{2\pi i(\theta+n\alpha+\frac{\alpha}{2})}-\lambda_3e^{-2\pi i(\theta+n\alpha^\prime+ \frac{\alpha^{\prime}}{2})}| \\
\leq &C(\lambda, N)\epsilon,
\end{align*}
and similarly, $|\tilde{c}_{\lambda}(\theta+m\alpha)-\tilde{c}_{\lambda}(\theta+m\alpha^\prime)| \leq C(\lambda, N)\epsilon$. Thus

\begin{align}\label{A-Aprime}
   \notag& \|A_{\lambda, \alpha}^E(\theta+m\alpha)-A^{E^\prime}_{\lambda, \alpha^\prime}(\theta+m\alpha)\| \\
 =\notag & \|\frac{D^E_{\lambda, \alpha}(\theta+m\alpha)}{c_{\lambda}(\theta+m\alpha)}-\frac{D^{E^\prime}_{\lambda, \alpha^\prime}(\theta+m\alpha)}{c_{\lambda}(\theta+m\alpha^\prime)}\| \\
\notag \leq & \frac{   \|D^E_{\lambda, \alpha}(\theta+m\alpha)\| |c_{\lambda}(\theta+m\alpha)-c_{\lambda}(\theta+m\alpha^\prime)|+|c_{\lambda}(\theta+m\alpha)|\|D^E_{\lambda,\alpha}(\theta+m\alpha)-D^{E^\prime}_{\lambda, \alpha^\prime}(\theta+m\alpha^\prime)\|}{|c_{\lambda}(\theta+m\alpha)c_{\lambda}(\theta+m\alpha^\prime)|}\\
\leq & C(\lambda, N, \alpha, \theta)\epsilon^{\frac{1}{2}}.
\end{align}

Take any normalized solution $u^{E^\prime}_{\ap, \theta}$ of $H_{\lambda, \alpha^\prime, \theta} u=E^\prime u$.
Take $u^{E}_{\alpha, \theta}$ be the solution of $H_{\lambda, \alpha, \theta}u=Eu$ with $u^E_{\alpha, \theta}(0)=u^{E^\prime}_{\alpha^\prime, \theta}(0)$ and $u^E_{\alpha, \theta}(-1)=u^{E^\prime}_{\alpha^\prime, \theta}(-1)$.

Then, by (\ref{A-Aprime}) and standard telescoping, we have that for any $|k|\leq N$,
\begin{align}\label{u-u}
\notag  &\|\left(\begin{matrix}u^{E\prime}_{\ap, \theta}(k)\\ u^{E\prime}_{\ap, \theta}(k-1)\end{matrix}\right)-\left(\begin{matrix}u^{E}_{\alpha, \theta}(k)\\ u^{E}_{\alpha, \theta}(k-1)\end{matrix}\right)\|\\
\notag =&\|\left(A^E_{\lambda,\alpha,k}(\theta)-A^{E^\prime}_{\lambda, \alpha^\prime, k}(\theta)\right)\left(\begin{matrix}u^{E}_{\alpha, \theta}(0)\\ u^{E}_{\alpha, \theta}(-1)\end{matrix}\right)\| \\ 
\notag \leq &\sum_{m=0}^{k}\| A^{E^\prime}_{\lambda, \ap, k-m-1}(\theta+(m+1)\ap)\left(A_{\lambda, \alpha}^{E}(\theta+m\alpha)-A_{\lambda, \alpha^\prime}^{E^\prime}(\theta+m\alpha^\prime) \right) A^E_{\lambda, \alpha, m}(\theta)\left(\begin{matrix}u^{E}_{\alpha, \theta}(0)\\ u^{E}_{\alpha, \theta}(-1)\end{matrix}\right)\|\\ 
\notag \leq &\sum_{m=0}^{k}\| A^{E^\prime}_{\lambda, \ap, k-m-1}(\theta+(m+1)\ap)\| \|\left(A_{\lambda, \alpha}^{E}(\theta+m\alpha)-A_{\lambda, \alpha^\prime}^{E^\prime}(\theta+m\alpha^\prime) \right)\| \|A^E_{\lambda, \alpha, m}(\theta)\|\\
\leq &C(\lambda, N, \alpha, \theta)\epsilon^{\frac{1}{2}}.
\end{align}
This implies 
\begin{equation}
(\sum_{|k|\leq N}|u^{E^{\prime}}_{\ap, \theta}(k)|^2)^{\frac{1}{2}} \geq (\sum_{|k|\leq N}|u^{E}_{\alpha, \theta}(k)|^2)^{\frac{1}{2}} -(\sum_{|k|\leq N}|u^{E^{\prime}}_{\ap, \theta}(k)-u^{E}_{\alpha, \theta}(k)|^2)^{\frac{1}{2}}
\geq C-C(\lambda, N, \alpha, \theta)\epsilon^{\frac{1}{2}}>C^\prime,
\end{equation}
provided $\epsilon<\epsilon (\lambda, \alpha, \theta, C-C^\prime, N)$.
$\hfill{} \Box$

Now we are ready to present the proof of the singular continuous part of Theorem \ref{main}. 
\subsection*{Proof of Theorem \ref{main}}

First, we describe the process briefly. Fix any $\alpha\in \{\alpha: \beta(\alpha)=L(\lambda)\}$, we are going to construct a convergent sequence of $\alpha_k\in \{\alpha: \beta(\alpha)=L(\lambda)+\frac{1}{2^k}\}$ such that $(\lambda, \alpha_k, \theta)$ is $(2k+3, N_k)$-bad. Then argue that $(\lambda, \alpha_{\infty}:=\lim_{k\rightarrow\infty}\alpha_k, \theta)$ is $(2k+2, N_k)$-bad for any $k\in \N$.

Now we will show the detailed construction step by step. 

Fix $\lambda$, $\theta\in \Theta$ and any $\alpha\in \{\alpha: \beta(\alpha)=L(\lambda)\}$. Let $\beta_k=L(\lambda)+\frac{1}{2^k}$. For any $\varepsilon>0$, our goal is to find $\alpha_{\infty}\in \{\alpha: \beta(\alpha)=L(\lambda)\}$ such that $|\alpha-\alpha_{\infty}|<\varepsilon$ and $H_{\lambda, \alpha_{\infty}, \theta}$ has purely singular continuous spectrum.
\subsection*{Step 0}
\

First, take $n_0\in \N$ large so that 
\begin{align}\label{n0choice}
\frac{1}{q_{n_0}^2(\alpha)}<\frac{1}{4}\varepsilon.
\end{align}

Take $\alpha_0$ to be defined as follows.
\begin{align}\label{alpha0}
a_n(\alpha_0)=
\left\{
\begin{matrix}
a_n(\alpha), \ \ &\mathrm{if}\ n\leq n_0,\\
[e^{\beta_0 q_{n-1}(\alpha_0)}],\ \ &\mathrm{if}\ n>n_0.
\end{matrix}
\right.
\end{align}
This choice of $\alpha_0$ guarantees the follows.
\begin{enumerate}
\item By (\ref{alpha0}), $p_{n_0}(\alpha_0)=p_{n_0}(\alpha)$ and $q_{n_0}(\alpha_0)=q_{n_0}(\alpha)$.
\item By (\ref{alpha-pnqn}) and (\ref{n0choice}), $|\alpha-\alpha_0|\leq |\alpha-\frac{p_{n_0}(\alpha)}{q_{n_0}(\alpha)}|+|\alpha_0-\frac{p_{n_0}(\alpha_0)}{q_{n_0}(\alpha_0}|<\frac{1}{q_n^2(\alpha)}+\frac{1}{q_n^2(\alpha_0)}<\frac{1}{4}\varepsilon+\frac{1}{4}\varepsilon=\frac{1}{2}\varepsilon$.
\item By (\ref{alpha0}), $\beta(\alpha_0)=\beta_0$.
\item By Lemma \ref{existCNbad}, there exists $N_0\in \N$ such that $(\lambda, \alpha_0, \theta)$ is $(3, N_0)$-bad.
\end{enumerate}

\subsection*{Step k, $k\geq 1$}
\

Given $\alpha_{k-1}$, choose $n_k\geq 2n_{k-1}$ such that
\begin{align*}
\frac{1}{q_{n_k}^2(\alpha_{k-1})}<\frac{1}{2}\min{(\frac{\varepsilon}{2^k}, \epsilon(\lambda, \alpha_{k-1}, \theta, 1, N_{k-1}))}.
\end{align*}
Now take $\alpha_k$ defined as follows.
\begin{align*}
a_n(\alpha_k)=
\left\{
\begin{matrix}
a_n(\alpha_{k-1}),\ \ &\mathrm{if}\ n\leq n_k,\\
[e^{\beta_k q_{n-1}(\alpha_k)}],\ \ &\mathrm{if}\ n>n_k.
\end{matrix}
\right.
\end{align*}
This choice of $\alpha_k$ guarantees the follows.
\begin{enumerate}
\item $p_{n_k}(\alpha_k)=p_{n_k}(\alpha_{k-1})$ and $q_{n_k}(\alpha_k)=q_{n_k}(\alpha_{k-1})$.
\item $|\alpha_{k-1}-\alpha_k|<\frac{\varepsilon}{2^{k+1}}+\frac{\varepsilon}{2^{k+1}}=\frac{\varepsilon}{2^k}$.
\item $\beta(\alpha_k)=\beta_k$.
\item By Lemma \ref{existCNbad}, there exists $N_k\in \N$ such that $N_k\geq 2N_{k-1}$ and $(\lambda, \alpha_k, \theta)$ is $(2k+3, N_k)$-bad.
\end{enumerate}

\subsection*{Step $\infty$, taking limit}
\

By our construction, $|\alpha_k-\alpha_{k-1}|<\frac{\varepsilon}{2^k}$, hence $\{\alpha_k\}$ is a Cauchy sequence. Let $\alpha_{\infty}=\lim_{k\rightarrow\infty}\alpha_k$, clearly $\alpha_{\infty}$ is a irrational number. 
Since the function $a_m(\alpha)$ is continuous at at irrational numbers, we have,
\begin{align}\label{alphainfty}
a_n(\alpha_{\infty})=a_n(\alpha_{m})\ \ \mathrm{for}\ n\leq n_k\ \mathrm{and}\ m\geq k-1.
\end{align}
Thus we have the following
\begin{enumerate}
\item $|\alpha_{\infty}-\alpha|\leq |\alpha-\alpha_0|+\sum_{m=1}^{\infty}|\alpha_m-\alpha_{m-1}|<\varepsilon$.
\item $\beta(\alpha_{\infty})=L(\lambda)$.
\item $|\alpha_{\infty}-\alpha_{k-1}|\leq |\alpha_{\infty}-\frac{p_{n_k}(\alpha_{\infty})}{q_{n_k}(\alpha_{\infty})}|+|\alpha_{k-1}-\frac{p_{n_k}(\alpha_{k-1})}{q_{n_k}(\alpha_{k-1})}|<\epsilon (\lambda, \alpha_{k-1}, \theta, 1, N_{k-1})$.
\end{enumerate}
In particular, by Lemma \ref{CNCprimeN}, part (3) and $(\lambda, \alpha_{k-1}, \theta)$ being $(2k+1, N_{k-1})$-bad imply $(\lambda, \alpha_{\infty}, \theta)$ being $(2k, N_{k-1})$-bad for any $k\in \N$. 
Purely singular continuous spectrum of $H_{\lambda, \alpha_{\infty}, \theta}$ then follows from Lemma \ref{CNpsc}. $\hfill{} \Box$

\section{Pure point spectrum}
Throughout this section, we will assume $\lambda\in \mathrm{I}^{\mathrm{o}}$.

Similar to \cite{AJZ}, the proof of pure point spectrum will be a corollary of the following reducibility result.
\begin{thm}\label{pp}
Let $0<\beta(\alpha)<\infty$. There exists a dense set of $\alpha$ with $\beta(\alpha)=L(\lambda)$ such that for a full measure set of $E\in \Sigma_{\sigma(\lambda), \alpha}$, the extended Harper cocycle $(\alpha, A^E_{\sigma(\lambda), \alpha})$ is $C^{\infty}$-reducible.
\end{thm}
\begin{rem}
The following Lemma \ref{AconjtildeA} reduces proving full measure reducibility of $(\alpha, A^E_{\sigma(\lambda), \alpha})$ to reducibility of $(\alpha, \tilde{A}^E_{\sigma(\lambda), \alpha})$.
\end{rem}

\begin{lemma}(see Appendix A of \cite{dryten})\label{AconjtildeA}
When $\beta(\alpha)\leq L(\lambda)$, there exists analytic $M(x)\in C^{\omega}(\T,GL(2,\mathbb C))$ such that
\begin{align*}
M(x+\alpha)\tilde{A}^E_{\sigma(\lambda), \alpha}(x)M^{-1}(x)=A^E_{\sigma(\lambda), \alpha}(x)
\end{align*}
namely, $(\alpha,\tilde{A}^E_{\sigma(\lambda), \alpha})$ and $(\alpha, A^E_{\sigma(\lambda), \alpha})$ are analytically conjugate to each other.
\end{lemma}

Our proof of Theorem \ref{pp} will rely on the following lemma. 

\begin{lemma}\label{reducible}
Let $\beta=L(\lambda)$, $0<\eta<\beta$, $\tau>1$ and $\gamma_1,\gamma_2>0$. 
Let $\alpha$ be an irrational number with $a_i(\alpha)\equiv 1$ for $i\geq N_0$.
For $n\geq N_0$, let $\alpha^{(n)}$ be defined as follow,
\begin{align}\label{ppalphaprime}
a_i(\alpha^{(n)})=
\left\lbrace
\begin{matrix}
a_i(\alpha),\ \ & i\leq n-1, \\
[e^{(\beta-\eta)q_{n-1}(\alpha)}],\ \ & i=n \\
1, &i\geq n+1
\end{matrix}
\right.
\end{align}
Suppose that $\rho(\alpha,\tilde{A}^E_{\sigma(\lambda), \alpha}) \in \mathrm{DC}_{\alpha}(\gamma_1,\tau)$ and
$\rho(\alpha^{(n)},\tilde{A}^E_{\sigma(\lambda), \alpha^{(n)}}) \in \mathrm{DC}_{\alpha^{(n)}}(\gamma_2,\tau)$.
Then there exists $B \in C^{\omega}(\T, \mathrm{SL}(2,\R))$ and $B^{(n)} \in C^{\infty}(\T, \mathrm{SL}(2,\R))$, such that
\begin{enumerate}
\item $(\alpha,\tilde{A}^E_{\sigma(\lambda), \alpha})$ is reducible by $B$ to $R_{\rho(\alpha, \tilde{A}^E_{\sil, \alpha})}$;
\item $(\alpha^{(n)},\tilde{A}^E_{\sigma(\lambda), \alpha^{(n)}})$ is reducible by $B^{(n)}$ to $R_{\rho(\alpha^{(n)}, \tilde{A}^E_{\sil, \alpha^{(n)}})}$.
\end{enumerate}
Moreover, for any $\varepsilon>0$, there exists $N=N(\alpha,\eta,\tau,\gamma_1,\gamma_2,\varepsilon)$ such that for $n\geq N$, we have
\begin{align*}
|\alpha-\alpha^{(n)}|<\varepsilon\ \ \mathrm{and}\ \ \mathrm{dist}_{C^{\infty}}(B, B^{(n)})<\varepsilon.
\end{align*}
\end{lemma}

\subsection*{Proof of Lemma \ref{reducible}}
\

This proof is based on a duality argument which leads to reducibility of $(\alpha, \tilde{A}^E_{\sil, \alpha})$ and then a perturbation argument that leads to reducibility of $(\alpha^{(n)}, \tilde{A}^E_{\sil, \alpha^{(n)}})$.

For simplicity, within this proof, we denote $p_k(\alpha), q_k(\alpha)$ by $p_k, q_k$ and $p_k(\an), q_k(\an)$ by $\tilde{p}_k, \tilde{q}_k$. We point out that by our construction, $p_k=\tilde{p}_k$ and $q_k=\tilde{q}_k$ for $1\leq k\leq n-1$.

With $\alpha\in \mathrm{DC}$ (see section \ref{seccontinued}) and $\rho(\alpha, \tilde{A}^E_{\sil, \alpha})$, the duality argument, thus reducibility of $(\alpha, \tilde{A}^E_{\sil, \alpha})$, is a quick combination of Theorem \ref{asymthm} with the following theorem.

\begin{thm}\label{dryten_reducible}\cite{dryten}
Let $\lambda\in \mathrm{I}^{\mathrm{o}}$, $\alpha\in \R\setminus \Q$ and $\theta\in \R$, fix $E\in \Sigma_{\sil, \alpha}$, and suppose $H_{\lambda, \alpha, \theta}u=\frac{E}{\lambda_2} u$ has a non-zero exponentially decaying eigenfunction $u={\{u_k\}}_{k\in\Z}$, $|u_k|\leq e^{-c |k|}$ for $k$ large enough. Then for any small enought $\epsilon>0$, the following hold:
\begin{itemize}
\item (A) If $2\theta\notin \alpha\Z+\Z$, then there exists $B\in C^{\omega}_{\frac{c-\epsilon}{2\pi}}(\T, \mathrm{SL}(2, \R))$, with $\deg{B}=0$, such that 
$$
{B(x+\alpha)}\tilde{A}^E_{\sil, \alpha}(x)B^{-1}(x)=R_{\rho(\alpha, \tilde{A}^E_{\sil, \alpha})}.
$$
\item (B) If $2\theta \in \alpha\Z+\Z$ and $\alpha\in\DC$, then there exists $B\in C^{\omega}_{\frac{c-\epsilon}{2\pi}}(\T, \mathrm{PSL}(2, \R))$, such that 
$$
{B(x+\alpha)}\tilde{A}^E_{\sil, \alpha}(x)B^{-1}(x)=\left(\begin{matrix}\pm 1       &a\\ 0      &\pm 1\end{matrix}\right)
$$
with $a\neq 0$. In this case $\rho(\alpha, \tilde{A}^E_{\sil, \alpha})=\frac{m}{2}\alpha$ $\mathrm{mod}\ \Z$, where $m=\deg{B}$.
\end{itemize}
\end{thm}

We point out that $\rho(\alpha, \tilde{A}^E_{\sil, \alpha})\in \mathrm{DC}_{\alpha}(\tau, \gamma_1)$ implies that $H_{\lambda, \alpha, \theta(E)}u=\frac{E}{\lambda_2}u$ has exponentially decaying eigenfunction for certain $\theta(E)\in \mathrm{DC}_{\alpha}(\tau)$, this is a deep corollary of Theorem 3.4 of \cite{dryten}. 
Then combining Theorem \ref{asymthm} with part (A) of Theorem \ref{dryten_reducible}, 
we get $B\in C^{\omega}_{\frac{L(\lambda)-\eta/6}{2\pi}}(\T, \mathrm{SL}(2, \R))$ and $T=T(\alpha, \eta, \gamma_1, \tau)$ such that,
\begin{equation}\label{R}
B(x+\alpha)\tilde{A}^E_{\sigma(\lambda), \alpha}(x)B^{-1}(x)=R_{\rho(\alpha,\tilde{A}^E_{\sigma(\lambda), \alpha})}
\end{equation}
$\deg B=0$ and 
\begin{equation}
\|B\|_{\frac{L(\lambda)-\eta/6}{2\pi}}\leq T
\end{equation}
By (\ref{R}) and the Cauchy estimate, we have
\begin{equation}\label{FR}
B(x+\alpha^{(n)})\tilde{A}^E_{\sil, \an}(x)B^{-1}(x)=e^{F(x)}R_{\rho(\alpha, \tilde{A}^E_{\sigma(\lambda), \alpha})},
\end{equation}
where
\begin{align*}
\|F\|_{\frac{L(\lambda)-\eta/3}{2\pi}}  
\leq &\|(B(x+\an)\tilde{A}^E_{\sigma(\lambda), \an}(x)-B(x+\alpha)\tilde{A}^E_{\sigma(\lambda), \alpha}(x))\|_{\frac{L(\lambda)-\eta/3}{2\pi}} \cdot \|B\|_{\frac{L(\lambda)-\eta/6}{2\pi}} \\
\leq &|\an-\alpha)|\cdot \|\partial B\|_{\frac{L(\lambda)-\eta/3}{2\pi}}\cdot \| \tilde{A}^E_{\sigma(\lambda), \an}\|_{\frac{L(\lambda)-\eta/3}{2\pi}}\cdot \|B\|_{\frac{L(\lambda)-\eta/6}{2\pi}} \\
&+|\an-\alpha|\cdot \|B\|_{\frac{L(\lambda)-\eta/6}{2\pi}}\cdot \|\partial \tilde{A}^E_{\sil, \alpha}\|_{\frac{L(\lambda)-\eta/3}{2\pi}}\cdot \| B\|_{\frac{L(\lambda)-\eta/6}{2\pi}} \\
\leq &\frac{C(\lambda)T^2}{\eta^2 q_{n-1}^2}
\end{align*}
Then there exists $\tilde{N}=\tilde{N}(\alpha, \eta,\gamma_1,\gamma_2, \tau)$ such that if $n\geq \tilde{N}$, we have 
\begin{equation}\label{F}
\|F\|_{\frac{L(\lambda)-\eta/3}{2\pi}} \leq \frac{C(\lambda)T^2}{\eta^2 q_{n-1}^2} \leq \varepsilon(\tau, \gamma_2, \frac{\eta}{12\pi}),
\end{equation} where $\varepsilon(\tau, \gamma, h-h^{\prime})$ is defined in Theorem \ref{KAM}.

Also we have
\begin{equation}\label{rhoDC}
\rho(\an, e^{F(x)}R_{\rho(\alpha,\tilde{A}^E_{\sigma(\lambda), \alpha})})
=\rho(\an,\tilde{A}^E_{\sigma(\lambda), \an})\in \mathrm{DC}_{\an}(\tau,\gamma_2),
\end{equation}
which follows quickly from $\rho(\an,\tilde{A}^E_{\sigma(\lambda), \an}) \in \mathrm{DC}_{\an}(\tau,\gamma_2)$, and $\deg B=0$.  
Combining Theorem \ref{KAM} with (\ref{F}) and (\ref{rhoDC}), we obtain that there exists 
$\tilde{B}^{(n)}_1\in C^{\omega}_{\frac{\lambda-\eta/2}{2\pi}}(\T, \mathrm{SL}(2,\R))$ such that 
\begin{equation}\label{Bn1redu}
\tilde{B}^{(n)}_1(x+\an)e^{F(x)}R_{\rho(\alpha,\tilde{A}^E_{\sigma(\lambda), \alpha})}(\tilde{B}^{(n)}_1(x))^{-1}=R_{\phi_n(x)}
\end{equation} 
with the following estimates:
\begin{equation}\label{Bn-Id}
\|\tilde{B}^{(n)}_1-\mathrm{Id}\|_{\frac{L(\lambda)-\eta/2}{2\pi}} \leq \|F\|_{\frac{L(\lambda)-\eta/3}{2\pi}}^{\frac{1}{2}} \leq \frac{C(\lambda)T}{\eta q_{n-1}},
\end{equation}
\begin{equation}\label{phinupperbd}
\|\phi_n(x)-\hat{\phi}_n(0)\|_{\frac{L(\lambda)-\eta/2}{2\pi}}\leq \frac{C(\lambda)T^2}{\eta^2 q_{n-1}^2}.
\end{equation}

Now let $\psi(x)$ be such that $\hat{\psi}(0)=0$ and
\begin{equation}\label{psi-psi=phi}
\psi_n(x+\an)-\psi_n(x)= \phi_n(x)-\hat{\phi}_n(0),
\end{equation}
then
\begin{equation}\label{hatpsin}
\hat{\psi}_n(k)=\frac{\hat{\phi}_n(k)}{e^{2\pi ik\an}-1},\ \ k\neq 0.
\end{equation}
We will now show such $\psi(x)\in C^{\infty}(\T, \R)$.

Recall that for $s\geq 0$, we have
\begin{align*}
\|f\|_{C^s(\T)}\leq \sum_{j=0}^s\sum_{k\in\Z}|k|^j|\hat{f}(k)|
\end{align*}
Hence by (\ref{phinupperbd})(\ref{hatpsin}) and continued fraction estimates, we have the following estimates:
\begin{align}\label{psiCs}
\notag&\frac{1}{s+1}\|\psi_n\|_{C^s(\T)}\\
\notag\leq & \sum_{k\in\Z} |k|^{s+1} \frac{|\hat{\phi}_n(k)|}{\|k\an\|_{\T}} \\
\notag\leq & \left(\sum_{0<|k|< q_{n-1}}+\sum_{q_{n-1}\leq |k|<\tilde{q}_n}+\sum_{|k|\geq \tilde{q}_n}\right)  \frac{|k|^{s+1} |\hat{\phi}_n(k)|}{ \|k\an\|_{\T}} \\
\notag\leq & \left(\sum_{0<|k|< q_{n-1}}+\sum_{q_{n-1}\leq |k|<\tilde{q}_n}+\sum_{|k|\geq \tilde{q}_n}\right)  \frac{|k|^{s+1} e^{-(L(\lambda)-\eta/2)|k|}}{\|k\an\|_{\T}}\|\phi_n-\hat{\phi}(0)\|_{\frac{L(\lambda)-\eta/2}{2\pi}} \\
\notag\leq &\ \ 2q_{n-1}\sum_{0<|k|< q_{n-1}}|k|^{s+1}e^{-(L(\lambda)-\eta/2)|k|} \|\phi_n-\hat{\phi}(0)\|_{\frac{L(\lambda)-\eta/2}{2\pi}} \\
\notag      & + \sum_{q_{n-1}\leq |k|<\tilde{q}_n} 2e^{(L(\lambda)-\eta)q_{n-1}}|k|^{s+2}e^{-(L(\lambda)-\eta/2)|k|} \|\phi_n-\hat{\phi}(0)\|_{\frac{L(\lambda)-\eta/2}{2\pi}} \\
\notag       & + \sum_{|k|\geq \tilde{q}_n} 4|k|^{s+2} e^{-(L(\lambda)-\eta/2)|k|} \|\phi_n-\hat{\phi}(0)\|_{\frac{L(\lambda)-\eta/2}{2\pi}}\\
\leq &\frac{C(s,\lambda)T^2}{\eta^2 q_{n-1}}, 
\end{align}
which implies $\psi_n \in C^{s}(\T,\R)$ for any $s\geq 0$, hence $\psi_n\in C^{\infty}(\T, \R)$. 

Let $B^{(n)}_2(x)=R_{-\psi_n(x)}\tilde{B}^{(n)}_1(x)\in C^{\infty}(\T, \mathrm{SL}(2, \R))$, by (\ref{Bn1redu}) and (\ref{psi-psi=phi}), we have
\begin{align}\label{Bn2redu}
B^{(n)}_2(x+\an)e^{F(x)}R_{\rho(\alpha, \tilde{A}^E_{\sil, \alpha})}(B^{(n)}_2(x))^{-1}=R_{\hat{\phi}(0)}.
\end{align}

Finally, let $B^{(n)}(x)=B^{(n)}_2(x)B(x)\in C^{\infty}(\T, \mathrm{SL}(2, \R))$, by (\ref{FR}) and (\ref{Bn2redu}), we have
\begin{align}\label{Bnredu}
B^{(n)}(x+\an)\tilde{A}^E_{\sil, \an}(x)(B^{(n)}(x))^{-1}=R_{\hat{\phi}(0)}.
\end{align}

By (\ref{Bn-Id}) and (\ref{psiCs}), we also have the following estimates:
\begin{align}\label{Bn-BCs}
\notag \|B^{(n)}-B\|_{C^s(\T)}
\notag =&\|R_{-\psi_n(x)}\tilde{B}^{(n)}_1(x)B(x)-B(x)\|_{C^s(\T)}\\
\notag\leq &\|R_{-\psi_n(x)}\tilde{B}^{(n)}_1(x)-R_{-\psi_n(x)}+R_{-\psi_n(x)}-\mathrm{Id}\|_{C^s(\T)} \cdot \|B(x)\|_{C^s(\T)} \\
\notag \leq &(\|R_{-\psi_n(x)}-\mathrm{Id}\|_{C^s(\T)} + \|\tilde{B}^{(n)}_1(x)-\mathrm{Id}\|_{\frac{L(\lambda)-\eta/2}{2\pi}}) \cdot T \\
\notag \leq & T \left(\frac{C(\lambda)T}{\eta q_{n-1}} +\frac{C(s,\lambda)T^2}{\eta^2 q_{n-1}} \right) \\
<&\frac{\epsilon}{2},
\end{align}
provided $n>N(\alpha, \eta, \tau, \gamma_1, \gamma_2, \epsilon, s)$.

Choosing $N=\max_{1\leq s\leq -\frac{2\ln{\epsilon}}{\ln{2}}}N(\alpha, \tau, \tau, \gamma_1, \gamma_2, \epsilon, s)$, then for $n>N$, we have 
\begin{align*}
\mathrm{dist}_{C^{\infty}}(B, B^{(n)})=\sum_{s=0}^{\infty}2^{-n}\frac{\|B-B^{(n)}\|_{C^s}}{1+\|B-B^{(n)}\|_{C^s}}<\epsilon.
\end{align*}

We also point out that since $\deg{B}=0$ and $\|B-B^{(n)}\|_{C^0}<\frac{\epsilon}{2}$, we have $\deg{B^{(n)}}=0$, therefore (\ref{Bnredu}) becomes
\begin{align*}
B^{(n)}(x+\an)\tilde{A}^E_{\sil, \an}(x)(B^{(n)}(x))^{-1}=R_{\rho(\an, \tilde{A}^E_{\sil, \an})}.
\end{align*}
$\hfill{} \Box$

\subsection*{Proof of Theorem \ref{pp}}
First, we describe the process briefly. Fix any $\alpha\in \{\alpha: \beta(\alpha)=L(\lambda)\}$, we are going to construct a convergent sequence of $\alpha_k\in \mathrm{DC}$ and a convergent matrix sequence $B_k\in C^{\infty}(\T, \mathrm{SL}(2, \R))$, such that $(\alpha_k, \tilde{A}^E_{\sigma(\lambda), \alpha_k})$ is reducible by $B_k$ for a nearly full measure set of $E\in \R$. 
Then argue that $(\alpha_{\infty}:=\lim_{k\rightarrow\infty}\alpha_k, \tilde{A}^E_{\sigma(\lambda), \alpha_{\infty}})$ is reducible by $B_{\infty}:=\lim_{k\rightarrow\infty}B_k$ for a full measure set of $E$, this is made possible through a Borel-Cantelli argument.

Now we will present the detail process step by step.

Fix any $\alpha\in \{\alpha: \beta(\alpha)=L(\lambda)\}$ and constants $\gamma>0$, $\tau>1$, 
$0<\eta<\min{(1, \beta(\alpha))}$. For any $\varepsilon>0$, our goal is to find $\alpha_{\infty}\in \{\alpha: \beta(\alpha)=L(\lambda)\}$ such that $|\alpha-\alpha_{\infty}|<\varepsilon$ and the normalized cocycle $(\alpha_{\infty}, \tilde{A}^E_{\sigma(\lambda), \alpha_{\infty}})$ is full measure reducible.
\subsection*{Step 0}
\
First, take $n_0\in \N$ large so that
\begin{align}\label{ppn0}
|\alpha-\frac{p_{n_0}(\alpha)}{q_{n_0}(\alpha)}|<\frac{1}{4}\varepsilon.
\end{align}

Take $\alpha_0$ to be defined as follow.
\begin{align}\label{ppalpha0}
\alpha_0=[a_1(\alpha), a_2(\alpha), \cdots, a_{n_0}(\alpha), 1, 1, \cdots]
\end{align}
This choice of $\alpha_0$ guarantees the follow.
\begin{enumerate}
\item $p_{n_0}(\alpha_0)=p_{n_0}(\alpha)$ and $q_{n_0}(\alpha_0)=q_{n_0}(\alpha)$.
\item $|\alpha-\alpha_0|<\frac{1}{2}\varepsilon$.
\item $\alpha_0\in \mathrm{DC}$.
\end{enumerate}

\subsection*{Step k, $k\geq 1$}
\

Applying Lemma \ref{reducible} to $\alpha_{k-1}$, we know that there exists $n_k:=N(\alpha_{k-1}, \frac{\eta}{2^k}, \tau, \frac{\gamma}{2^{k-1}}, \frac{\gamma}{2^k}, \frac{\varepsilon}{2^{k+1}})$, such that the two frequencies $\alpha_{k-1}$ and $\alpha_k:=\alpha_{k-1}^{(n_k)}$ satisfy the following properties.
\begin{enumerate}
\item $\alpha_{k}\in \mathrm{DC}$.
\item $|\alpha_{k-1}-\alpha_k|<\frac{\varepsilon}{2^{k+1}}$.
\item $a_{n_k}(\alpha_k)=[e^{(L(\lambda)-\frac{\eta}{2^k})q_{n_k-1}(\alpha_k)}]$.
\item if $\rho(\alpha_{k-1}, \tilde{A}^E_{\sil, \alpha_{k-1}})\in \mathrm{DC}_{\alpha_{k-1}}(\tau, \frac{\gamma}{2^{k-1}})$ and 
$\rho(\alpha_k, \tilde{A}^E_{\sil, \alpha_k})\in \mathrm{DC}_{\alpha_k}(\tau, \frac{\gamma}{2^k})$, 
then there exist $B_{k-1}^{\omega}\in C^{\omega}(\T, \mathrm{SL}(2, \R))$ and 
$B_k^{\infty}\in C^{\infty}(\T, \mathrm{SL}(2, \R))$ 
such that $(\alpha_{k-1}, \tilde{A}^E_{\sil, \alpha_{k-1}})$ is reducible to $R_{\rho(\alpha_{k-1}, \tilde{A}^E_{\sil, \alpha_{k-1}})}$ by $B_{k-1}^{\omega}$, and $(\alpha_k, \tilde{A}^E_{\sil, \alpha_k})$ is reducible by $B_k^{\infty}$ to $R_{(\alpha_k, \tilde{A}^E_{\sil, \alpha_k})}$.
\item $\mathrm{dist}_{C^{\infty}}(B_{k-1}^{\omega}, B_k^{\infty})<\frac{\varepsilon}{2^{k+1}}$.
\end{enumerate}

\subsection*{Step $\infty$, taking limit}
\

First we recall the following lemma whose proof is a combination of full measure rotations reducibility \cite{AFK, YouZhou} and formula (1.5) of \cite{AFK}.
\begin{lemma}\label{measure}
Let $\lambda\in \mathrm{I}^{\mathrm{o}}$, $\alpha\in \R\setminus \Q$. 
For a full measure set of $E\in \Sigma_{\sil, \alpha}$, there exists $B_E\in C^{\omega}(\T, \mathrm{SL}(2,\R))$ such that
\begin{equation}
B_E(\theta+\alpha)\tilde{A}^E_{\sil, \alpha}(\theta)B_E(\theta)^{-1}\in \mathrm{SO}(2,\R).
\end{equation}
Furthermore, we have
\begin{equation}
\frac{\mathrm{d} {\rho}(\alpha, \tilde{A}^E_{\sil, \alpha})}{\mathrm{d} E}=-\frac{1}{8\pi}\int_{\T}\|B_E(\theta)^2\|_{\mathrm{HS}}\ \mathrm{d}\theta \leq -\frac{1}{4\pi}
\end{equation}
Here $\|\cdot\|_{\mathrm{HS}}$ denotes the Hilbert-Schmidt norm.
\end{lemma}

As a direct corollary of the lemma, we have 
\begin{align}\label{measure2k}
  \notag&|\{E\in \R: \rho(\alpha_k, \tilde{A}^E_{\sil, \alpha_k})\notin \mathrm{DC}_{\alpha_k}(\tau, \frac{\gamma}{2^k})\}|\\
=\notag&|\{E\in \Sigma_{\sil, \alpha_k} \rho(\alpha_k, \tilde{A}^E_{\sil, \alpha_k})\notin \mathrm{DC}_{\alpha_k}(\tau, \frac{\gamma}{2^k})\}|\\
=&O(\frac{\gamma}{2^k}).
\end{align}

Take
\begin{align*}
\mathcal{B}=\cap_{n=1}^{\infty}\cup_{k= n}^{\infty}\{E\in \R: \rho(\alpha_k, \tilde{A}^E_{\sil, \alpha_k})\notin \mathrm{DC}_{\alpha_k}(\tau, \frac{\gamma}{2^k})\}.
\end{align*}
Then by Borel-Cantelli lemma, we have $|\mathcal{B}|=0$. Which implies for a.e. $E\in \R$, there exists $n_E\in \N$, such that $\rho(\alpha_k, \tilde{A}^E_{\sil, \alpha_k})\in \mathrm{DC}_{\alpha_k}(\tau, \frac{\gamma}{2^k})$ for any $k\geq n_E$.
By our construction of $\alpha_k$, this implies that for $k\geq n_E+1$, 
\begin{enumerate}
\item $(\alpha_k, \tilde{A}^E_{\sil, \alpha_k})$ is reducible by both $B^{\omega}_k\in C^{\omega}(\T, \mathrm{SL}(2, \R))$ and $B^{\infty}_k\in C^{\infty}(\T, \mathrm{SL}(2, \R))$ to $R_{(\alpha_k, \tilde{A}^E_{\sil, \alpha_k})}$.
\item $\mathrm{dist}_{C^{\infty}}(B^{\omega}_k, B^{\infty}_{k+1})<\frac{\varepsilon}{2^{k+2}}$.
\end{enumerate}

The following lemma will be proved in Appendix \ref{tworedu}:
\begin{lemma}\label{tworeducible}
Let $\rho(\alpha, \tilde{A}^E_{\sil, \alpha})\in \mathrm{DC}_{\alpha}(\tau, \gamma)$. Suppose there exists two matrix functions $B_1, B_2\in C^{\infty}(\T, \mathrm{SL}(2, \R))$ such that $(\alpha, \tilde{A}^E_{\sil, \alpha})$ is reducible by both of them to $R_{\rho(\alpha, \tilde{A}^E_{\sil, \alpha})}$. Then there exists a constant rotation $R_{\psi}$ so that $B_1(\theta)\equiv R_{\psi}B_2(\theta)$.
\end{lemma}

By Lemma \ref{tworeducible}, there exists $\psi_k\in [0, 1)$ such that $B^{\omega}_k=R_{\psi_k}B^{\infty}_k$. 
Hence, letting $B_{n_E+1}:=B_{n_E+1}^{\infty}$ and $B_k:=R_{\sum_{j=n_E+1}^k \psi_j} B_k^{\infty}$ for $k>n_E+1$, item (2) transforms into the following form.
\begin{align*}
\mathrm{dist}_{C^{\infty}}(B_k, B_{k+1})<\frac{\varepsilon}{2^{k+2}},\ \ \mathrm{for}\ k\geq n_E+1.
\end{align*}
Hence $\{\alpha_k\}$ and $\{B_k\}$ are both Cauchy sequences.

Taking $\alpha_{\infty}=\lim_{k\rightarrow\infty}\alpha_k$, by our construction,
\begin{enumerate}
\item $|\alpha-\alpha_{\infty}|\leq |\alpha-\alpha_0|+\sum_{k=1}^{\infty}|\alpha_k-\alpha_{k-1}|<\varepsilon.$
\item $a_{n_k}(\alpha_{\infty})=[e^{(L(\lambda)-\frac{\eta}{2^k})q_{n_k-1}(\alpha_{\infty})}]$, hence by (\ref{beta=limsupa}), $\beta(\alpha_{\infty})=\limsup_{k\rightarrow\infty}\frac{\ln{a_{n_k}(\alpha_{\infty})}}{q_{n_k-1}(\alpha_{\infty})}= L(\lambda)$.
\end{enumerate}
Finally, letting $B_{\infty}=\lim_{k\rightarrow\infty}B_k$, we conclude that $(\alpha_{\infty}, \tilde{A}^E_{\sil, \alpha_{\infty}})$ is reducible by $B_{\infty}$. $\hfill{} \Box$.

\appendix

\section{Proof of Lemma \ref{tworeducible}}\label{tworedu}
For simplicity, we will denote $\tilde{A}^E_{\sil, \alpha}$ by $A$.

Let $B_3=B_2B_1^{-1}\in C^{\infty}(\T, \mathrm{SL}(2, \R))$, then
\begin{align*}
B_3(\theta+\alpha)R_{\rho(\alpha, A)}=R_{\rho(\alpha, A)}B_3(\theta).
\end{align*}
This is equivalent to 
\begin{align}\label{App1}
B_4(\theta+\alpha)\left(\begin{matrix}e^{i\rho(\alpha, A)}\ &0\\ 0\ &e^{-i\rho(\alpha, A)}\end{matrix}\right)
=\left(\begin{matrix}e^{i\rho(\alpha, A)}\ &0\\ 0\ &e^{-i\rho(\alpha, A)}\end{matrix}\right) B_4(\theta),
\end{align}
with $B_4(\theta)=\left(\begin{matrix}i\ &-i\\ 1\ &1\end{matrix}\right)^{-1}B_3(\theta)
\left(\begin{matrix}i\ &-i\\ 1\ &1\end{matrix}\right)$.
Let $B_4(\theta)=\left(\begin{matrix}b_1(\theta)\ &b_2(\theta)\\ b_3(\theta)\ &b_4(\theta)\end{matrix}\right)$ and expanding (\ref{App1}), we get
\begin{align}
b_1(\theta+\alpha)&=b_1(\theta),\\
b_2(\theta+\alpha)&=e^{2i\rho(\alpha, A)}b_2(\theta),\\
b_3(\theta+\alpha)&=e^{-2i\rho(\alpha, A)}b_3(\theta),\\
b_4(\theta+\alpha)&=b_4(\theta).
\end{align}
(A.2) and (A.5) directly imply $b_1(\theta)\equiv b_1$, $b_4(\theta)\equiv b_4$ being constants. 
Taking Fourier coefficients, (A.3) and (A.4) imply the following:
\begin{align}
e^{-ik\alpha}\hat{b}_2(k)&=e^{2i\rho(\alpha, A)}\hat{b}_2(k),\ \ \forall k\in \Z.\\
e^{-ik\alpha}\hat{b}_3(k)&=e^{-2i\rho(\alpha, A)}\hat{b}_3(k),\ \ \forall k\in \Z.
\end{align}
Clearly, $\rho(\alpha, A)\in \mathrm{DC}_{\alpha}(\tau, \gamma)$ implies $\hat{b}_2(k)=\hat{b}_3(k)=0$ for any $k\in \Z$, hence $b_2(\theta)\equiv b_3(\theta)\equiv 0$.

In summary, $B_4(\theta)=\left(\begin{matrix}b_1\ &0\\ 0\ &b_1^{-1}\end{matrix}\right)$ since $\det{B_4}\equiv 1$. Taking into account that 
\begin{align*}
B_3=\left(\begin{matrix}i\ &-i\\ 1\ &1\end{matrix}\right)\left(\begin{matrix}b_1\ &0\\ 0\ &b_1^{-1}\end{matrix}\right) \left(\begin{matrix}i\ &-i\\ 1\ &1\end{matrix}\right)^{-1}\in \mathrm{SL}(2, \R),
\end{align*} 
we must have $b_1=e^{i\psi}$. 
Thus $B_3(\theta)=B_2(\theta)B_1^{-1}(\theta)\equiv R_{\psi}$.
This means $B_2(\theta)=R_{\psi}B_1(\theta)$. $\hfill{} \Box$
 
\section*{acknowledgement}

I would like to thank Qi Zhou for suggesting this problem. I also grateful thank Rui Han and Professor Svetlana Jitomirskaya for many useful discussions and comments. This research was supported by the NSF DMS-1401204 and NSF DMS-1500703.

\bibliographystyle{amsplain}

\end{document}